# Unusual Phase Behavior of Confined Heavy Water

Yang Zhang (zhyang@ornl.gov)[1,2], Antonio Faraone[3,4], William A. Kamitakahara[3], Kao-Hsiang Liu[5], Chung-Yuan Mou[5], Juscelino B. Leão[3], Sung Chang[3], Sow-Hsin Chen (sowhsin@mit.edu)[1]

[1]Department of Nuclear Science and Engineering, Massachusetts Institute of Technology, Cambridge, Massachusetts 02139, USA.

[2]Neutron Scattering Science Division and Joint Institute for Neutron Sciences, Oak Ridge National Laboratory, Oak Ridge, Tennessee 37831, USA.

[3]NIST Center for Neutron Research, National Institute of Standards and Technology, Gaithersburg, Maryland 20899, USA.

[4]Department of Materials Science and Engineering, University of Maryland, College Park, Maryland 20742, USA.

[5]Department of Chemistry, National Taiwan University, Taipei 106, Taiwan.

**Abstract**

Many of the anomalous properties of water are amplified in the deeply supercooled region. Here we present neutron scattering measurements of the density of heavy water confined in a nanoporous silica matrix MCM-41-S (≈15 Å pore diameter), namely, the equation of state $\rho(T,P)$, in a temperature-pressure range, from 300 K to 130 K and from 1 bar to 2900 bar, where bulk water will crystallize. A sudden change of slope in the otherwise continuous density profile (a "kink") is observed below a certain pressure $P_c$; however, this feature is absent above $P_c$. Instead, a hysteresis phenomenon in the density profiles between the warming and cooling scans becomes prominent above $P_c$. Hence, the data can be interpreted as a line of apparent 2nd-order phase transition at low pressures evolving into a line of 1st-order phase transition at high pressures. If so, the existence of a "tricritical point" at $P_c \approx 1500$ bar, $T_c \approx 210$ K becomes another possible scenario to explain the exceptionally rich phase behavior of low-temperature confined water.

**Introduction**

In many biological or geological systems, water resides in pores of nanoscopic dimensions, or close to hydrophilic or hydrophobic surfaces, comprising a layer of water, one or two molecules thick, with properties often different from the bulk. Such "confined" or "interfacial" water has attracted considerable attention, due to its fundamental importance in many processes, such as protein folding, concrete curing, corrosion, molecular and ionic transport, etc [1-3]. However, our understanding of the numerous physicochemical anomalies of confined water, and indeed of bulk water, is still incomplete. Basic gaps persist, among which the most interesting one is the origin of the unusual behavior of water in the supercooled region where water remains in the liquid state below the melting point [4-7]. Recent studies have aimed at explaining anomalies such as the density maximum and minimum [8-10], and the apparent divergence of the thermodynamic response functions at 228 K at ambient pressure [11]. The three major hypothesized scenarios currently under scrutiny are: the "singularity-free (SF) scenario" [12,13], the "liquid-liquid critical point (LLCP) scenario" [14,15] and the "critical point-free (CPF) scenario" [16]. It is hypothesized, by all these three scenarios, that in the low temperature range bulk water is composed of a mixture of two structurally distinct liquids: the low-density liquid (LDL) and the high-density liquid (HDL). They are respectively the thermodynamic continuation of the low-density amorphous ice (LDA) and high-density amorphous ice (HDA) into the liquid state. Evidence of a first-order phase transition between LDA and HDA has been reported since 1985 [17-20]. Subsequently, several experimental findings have been interpreted as support of the hypothetical existence of two different structural motifs of liquid water [21-27]. However, some of the interpretations have been questioned [28,29]. So far, direct evidence of a first-order liquid-liquid phase transition between LDL and HDL, as a thermodynamic extension of the first-order transition established in the amorphous solid waters, has not yet been observed.

An experimental challenge arises because the hypothesized first-order liquid-liquid phase transition exists in a region of the phase diagram, often called the "no man's land" [5], in which bulk water cannot exist in a liquid state. One method of overcoming this difficulty is to take advantage of confinement. By confining water in a nanoporous silica matrix, MCM-41-S with 15 Å pore diameter, the homogeneous nucleation process (crystallization) can be avoided, allowing us to enter the "no man's land" and investigate the properties of liquid water. There is still much debate on the differences and similarities between bulk and confined water [16,30-32], however, even if the silica matrix, with its hydrophilic surfaces, might affect properties of water other than the suppression of homogeneous nucleation, confined water in MCM-41-S is representative of many environments of interest in biological and geological sciences where similar hydrophilic interfaces are intrinsic and important [1].

In this paper, we describe an efficient method for the density determination employing the cold neutron triple-axis spectrometer SPINS at the NIST Center for Neutron Research (NCNR). Using this method, we are able to obtain sensitive measurements of the density of $D_2O$ confined in MCM-41-S as a function of closely spaced temperatures (1 K interval) from 300 K to 130 K in a range of pressures from ambient to 2900 bar, achieving a remarkably good signal-to-noise ratio. The reliability and accuracy of the method are extensively discussed in the Materials and Methods section. There, we also estimate the extent of possible effects related to small amounts of helium from the pressurizing system being dissolved in the water, and the layering water distributions along the pore radius direction, showing that such effects are most likely negligible. Density plays a central role in many classical phase transitions. In particular, it is the order parameter in the gas-liquid and liquid-solid transitions. Therefore, its experimental determination assumes primary importance regarding the hypothesized liquid-liquid phase transition. In making such measurements, we are seeking evidence of a remnant of a first-order liquid-liquid phase transition of water that would exist in the macroscopic system if it were possible to avoid crystallization.

**Results and Discussions**

As described in the Materials and Methods section, the contrast of the neutron coherent scattering length density (SLD) of heavy water against that of the silica matrix gives rise to a strong signal in our experiment. Specifically, we observe a well-defined first Bragg diffraction peak arising from the (10) plane of a 2-D hexagonal lattice of water cylinders in the grains of MCM-41-S silica matrix (Fig.1(a)). Fig.1(b-d) illustrate the elastic neutron diffraction intensities measured at the highest and lowest temperatures at three representative pressures. One can immediately notice that the width and the position of the Bragg peak do not change with temperature. Hence, for our purposes, the structure of the confining matrix can be regarded as unaffected by temperature. Once the data are corrected for the temperature independent background arising from the fractal packing of the MCM-41-S crystallites (grains) and the incoherent scattering, the only temperature-dependent quantity is the height of the Bragg peak, which is proportional to the square of the difference of SLD between the heavy water and the silica matrix, and therefore a sensitive indicator of the average mass density of the confined water. We can therefore sit at the Bragg peak position and monitor the peak intensity as a function of temperature, rather than performing a scan in Q at each temperature. While our measurements are highly precise and sensitive with regard to relative changes, there is an overall uncertainty that we estimate (from the results of repeated measurements on the same and different sample batches) to be about 0.02 g/cm$^3$ (standard deviation) in the overall density scale, arising from uncertainties in the scattering length density of the silica matrix, and the model we have used to analyze the data. Even after careful considerations of all these sources of uncertainties, the estimation of the uncertainty of the absolute density is still a challenge because of the possible systematic errors arising from the model used in the analysis. It should be pointed out, however, that this uncertainty can be considered as a scaling factor, and that the relative shape of the density curves is almost directly related to the measured scattering intensity.

The temperature dependence of the density of confined heavy water measured by cooling scans at pressures from ambient to 2900 bar are shown in Fig.2(a). The isobaric density profiles show a steep decrease as the water is increasingly cooled, reaching a clear minimum for each pressure. The minimum temperature $T_{min}$ decreases from 210 K to 170 K as the pressure is increased from ambient to 2900 bar. Poole et al. have proposed that the occurrence of such a density minimum is an indication of full development of a defect-free random tetrahedral network (RTN) of the hydrogen bonds [33]. Below $T_{min}$ the completed RTN shows normal thermal contraction as the temperature is further lowered. Our results therefore imply that at higher pressures, the RTN can only be reached at lower temperatures. This is a consequence of the fact that the enthalpically favourable hydrogen bonded RTN has a lower density compared to its less developed counterpart.

Another remarkable feature of this plot is that the isobaric density profile shows a sudden change of slope ("kink") at a certain temperature at and below ≈1600 bar. In order to understand this phenomenon, we fitted the density profiles with 2-piecewise 3$^{rd}$-order polynomial at pressures from ambient to 2000 bar. The isobaric thermal expansion coefficients $α_P$ are then evaluated analytically from the fitted polynomials, shown in Fig. 2(b). A discontinuity in $α_P$ directly coming from the "kink" is presented at and below ≈1600 bar. A discontinuity in the response functions defines a 2$^{nd}$-order phase transition [34]. Whether the observed transition belongs to the $λ$ transition universality class requires more accurate studies of the critical exponents. The most prominent example of a 2$^{nd}$-order phase transition is the superfluid transition of helium-4, which shows an analogous "kink" in the density profile [35] and a discontinuity in $α_P$ [35] and $c_P$ [36]. The possible existence of a $λ$-transition in bulk and confined low-temperature water was argued previously in the literature based on specific heat measurements [16], although the discontinuity in $c_P$ has never been observed experimentally due to finite temperature increment (instead, it is smeared out into a peak) [37]. Mean field theory and Monte Carlo simulations have also provided compelling evidences of the "kink" in the density profile [38] and the discontinuity in $c_P$ [39], while they were attributed to the crossing of the "Widom line" (a line of the maximum correlation length in the one phase region according to the LLCP scenario) [38,40]. Note that the growth of the correlation length $ξ$ (the peak height of $α_P$) on approaching 1600 bar from ambient pressure is slightly observable, probably because the growth of the correlation length $ξ$ is limited in the confined geometry.

Above ≈1600 bar, the "kink" in the density profile is no longer observable. Accordingly, the discontinuity in $α_P$ also disappears, implying an endpoint of the 2$^{nd}$-order phase transition line. Note that, at 2000 bar, the discontinuity in $α_P$ becomes minimal, independent of the enforced fitting with the piecewise polynomial. Therefore, at 2500 and 2900 bar, the density profiles are fitted with single polynomials (9$^{th}$-order) in order to take derivatives.

Normally, a discontinuous change of the state functions, such as density, associated with a first-order transition is difficult to detect directly. When the equilibrium phase boundary is crossed, due to the metastability or the kinetics of the phase transition, the phase separation may take very long time to happen, especially in confinement [41-43]. However, a first-order line should still manifest itself with a significant hysteresis when it is crossed from opposite directions of the transition line, shown in Fig.1(b). Although a hysteresis phenomenon, if it exists, may not prove the existence of a first-order transition, the absence of hysteresis can serve to rule out a first-order phase change. Accordingly, we performed a series of warming and cooling scans over a range of pressures. For each pressure, the sample was cooled from 300 K to 130 K at ambient pressure, and then pressurized to the desired value. We then waited about 2 hours for the system to equilibrate. The warming scan with 0.2 K/min was first performed from 130 K to 300 K. When the warming scan was finished, we waited another 2 hours at room temperature for system equilibration. After that, the cooling scan with 0.2 K/min was performed from 300 K to 130 K. When the full cycle was finished, the sample was brought back to ambient temperature and pressure before measuring another pressure. The system we choose to study is water confined in long cylindrical pores of silica with a diameter of 15 Å. In such confined water, two of its three dimensions are finite, leaving one along the pore axis that can be considered macroscopic. Thus, it is intrinsically difficult to observe a phase transition in such a restricted geometry. However, we may still find a remnant of a phase transition in a confined system, which in principle could be demonstrated by its size scaling.

In order to understand the change of behaviors in density and expansion coefficient at around 1600 bar, we performed density measurements with both cooling and warming scans at a series of pressures, shown in Fig.3(a). The difference between the two scans are shown in Fig.3(b). The fact that the warming and cooling curves join at both high and low temperature ends implies that the expansion-contraction processes are reversible. Up to 1000 bar, the density difference between the cooling and warming scans is small, which could be attributed to the temperature lag when ramping the temperature continuously. The density difference due to this reason is small and relatively independent of pressure. However above 1000 bar, the density difference (hysteresis) opens up progressively as the pressure is increased. We expect the magnitude of the density difference might depend on the temperature ramping rate. Considering the feasibility of neutron scattering experiments, we chose a ramping rate of 0.2 K/min. With such a slow rate, a rather uniform temperature distribution over the sample is assured, but the system may require much longer times to reach physical equilibrium after crossing a phase boundary. There seems to be a small regression of the opening up of the density difference at 2900 bar between 190 K and 210 K. Further investigations are needed to find out whether this regression is real or an experimental artifact. Nevertheless, the biggest density difference (confined $D_2O$) is found to be on the order of a few percent above 1000 bar. In comparison to the density difference between high and low density amorphous $H_2O$ ice (about 25%, measured at much lower temperatures), the observed difference is small. The reason, to our best speculation, might be a combined effect of confinement, isotopic difference and temperatures. Note that the accuracy of the absolute density we determined depends on the background subtraction and the scaling. However, the relative shape of the density profiles is independent of the analysis.

We now consider whether the observed density hysteresis can be related to a liquid-glass transition of confined heavy water [44-46], as distinct from a glass transition, in which the macroscopic observables may depend on the thermal history of the system. In the literature, the glass transition temperature of bulk $H_2O$ at ambient pressure is commonly accepted to be around 130 K [47,48], and is suggested to be modified to be about 160 K [49,50]. It is expected to be even lower at elevated pressures. Moreover, the structural relaxation time of the confined $H_2O$ was reported to be in the order of a few nanoseconds at around 220 K at ambient pressure [51,52] and even faster at elevated pressures [53]. Note that in our experiments we scan at 0.2 K/min, which is many orders of magnitude slower than the structural relaxation time of the confined water. Therefore, it is apparent that the maximum hysteresis we observe in confined $D_2O$ at high pressures happens far above the glass transition.

Finally, Fig.4 summarizes the lines of density "kinks" and maximum density hysteresis in the P-T plane. Our density measurements suggest that there is a 2$^{nd}$-order phase transition line at low pressures, and a 1$^{st}$-order phase transition line at high pressures. Therefore, in the moderate pressure range, there might exist a point that connects the 2$^{nd}$-order and 1$^{st}$-order lines. The term "tricritical point" is used in the literature for such a point [34,54,55]. A tricritical point is defined in parallel with an ordinary critical point, except that it is a point in the phase diagram at which three-phase (instead of two-phase) coexistence terminates [56]. The terminology comes from the view of the three dimensional phase space when an additional conjugate field of the order parameter is introduced. Well-known examples include, but are not limited to, the mixture of $^3$He and $^4$He [57,58], and metamagnet $FeCl_2$ [59].

For heavy water confined in MCM-41-S, our density measurements clearly show a change of behavior at $T_c = 210 \pm 10$ K, $P_c = 1500 \pm 500$ bar. Above $P_c$, the hysteresis phenomenon is a strong evidence of the existence of the 1$^{st}$-order phase transition, which is predicted for bulk water by the LLCP scenario. Below $P_c$, the discontinuity of the expansion coefficient suggests a 2$^{nd}$-order phase transition. Therefore, it is plausible that the hypothesized liquid-liquid critical point of water is actually a "tricritical point".

**Conclusions**

In summary, by using a triple-axis spectrometer, we were able to obtain sensitive and precise measurements of the density of confined heavy water, achieving a remarkably good signal-to-noise ratio. From the extracted density at closely spaced temperature intervals, we then evaluated the isobaric thermal expansion coefficient $\alpha_P$. We observe the occurrence of a sudden change of slope in the density $vs$ T profile and of a corresponding discontinuity in $\alpha_P$ at low pressures, which are interpreted as signatures of a 2$^{nd}$-order phase transition. At high pressures, we observe a hysteresis phenomenon, which is interpreted as a signature of a 1$^{st}$-order phase transition. On this basis, we speculate that a "tricritical point" might exist at around $T_c \approx 210 \pm 10$ K, $P_c \approx 1500 \pm 500$ bar for confined heavy water.

These findings give a unified framework for understanding the properties of confined water. The density data are the main results of the present paper and will be useful for a better modelling of the properties of water in various biological and geological conditions i) in rocks and clays, relevant for mining purposes and environmental questions; ii) on the surface of proteins and membranes, relevant to address at molecular level a number of biological processes; iii) in different artificial porous environments used for catalytic purposes.

**Materials and Methods**

**Sample preparation**. The MCM-41-S powder sample is made of micellar templated nanoporous silica matrices [60], consisting of grains of the order of micrometer size. In each grain, parallel cylindrical pores are arranged in a 2-D hexagonal lattice with an inter-plane distance $d = 30 \pm 2$ Å. The MCM-41-S is synthesized by reacting pre-formed $\beta$-zeolite seeds (composed by tetraethylammonium hydroxide (TEAOH, Acros), sodium hydroxide (NaOH) and fumed silica (Sigma)) with decyltrimethylammonium bromide solution (C$_{10}$TAB, Acros), then transferring the mixture into an autoclave at 120ºC for 2 days. After cooling down to room temperature, the mixture was adjust to pH = 10. Then the mixture was sealed into autoclave at 100 ºC for 2 days. Solid sample is collected by filtration, washed by water, dried at 60 ºC in air overnight, and then calcined at 540 ºC for 8 hours. The molar ratios of the reactants are SiO$_2$ : NaOH : TEAOH : C$_{10}$TAB : H$_2$O = 1 : 0.075 : 0.285 : 0.204 : 226.46. The pore diameter and pore volume are estimated to be $15 \pm 2$ Å and 0.50 cm$^3$/g respectively with the Barret-Joyner-Halenda (BJH) analysis. The pore diameter is also confirmed by fitting the elastic diffraction profile.

The dry MCM-41-S sample is then hydrated by exposing to water vapor in a closed container at room temperature. The achieved full hydration level for D$_2$O corresponds to a fractional mass gain of 0.5 (mass of absorbed water/mass of dry MCM-41-S). We performed an experiment of the surface functionalization of the Si-OH in MCM-41-S to determine the surface functional group Si-O-Si-(CH$_3$)$_3$ to the saturation monolayer level. The surface density of Si-OH is determined to be 1.16 groups/nm$^2$ through the chemical analysis of the carbon content. After the subtraction of the Si-OH on the external surface of the MCM-41-S grain ($\approx$10%), we estimated H$_{silanol}$ / H$_{water}$ = 0.065. Since the Si-O-H bonds are very strong and are attached on rigid surfaces, the migration of silanol groups on the surface is not likely. Therefore, in the temperature range we worked on, we believe that the interaction between waters and surface Si-OH represent a contribution that is "constant" and water-like. That is, the hydrophilic silanol surface provides a small and constant perturbation to the confined water. It is known that near the hydrophilic surface such as silica, there is a layer of denser water, while in the center of the pores water distributes uniformly [32,61-67]. The behavior of water near a hydrophobic surface may be different because of the lack of compensating hydrogen bonds from the surface, and therefore requires more careful investigations [10,64-66,68]. In the pores of MCM-41-S, the long-range ice-like order cannot develop; thus, the homogeneous nucleation process is inhibited. DSC check was routinely performed to make sure of: (a) No freezing of bulk water. (b) No freezing of confined water occurs down to 130 K.

**Neutron scattering experiment**. In this experiment, we attempt to measure the average density of $D_2O$ confined in the pores of MCM-41-S. The measurement was carried out at the NIST Center for Neutron Research (NCNR) using the cold neutron triple-axis spectrometer SPINS, operated in an elastic scattering mode with incident neutron energy of 3.7 meV. The $D_2O$ hydrated MCM-41-S sample was loaded in the NCNR pressure cell HW-02 with a sample volume of 1.5 cm$^3$. $D_2O$ has a considerably different coherent neutron scattering length density from that of the silica matrix, giving rise to a well-defined Bragg peak. Pressure was applied with helium gas. The sample temperature was controlled using a top loading closed cycle refrigerator. A small amount of helium was used to insure thermal exchange between the sample and the wall of the refrigerator, whose temperature was controlled with accuracy better than 0.01 K. The density data are reported as function of the sample temperature, which is recorded by a sensor located just above the pressure cell.

The diffraction pattern of our sample consists of three parts: (i) the low-Q scattering of the fractal packing of the grains, which follows a power law Q-dependence; (ii) a Bragg peak at around *2π/d* coming from the 2-D hexagonal internal structure of the grains; and (iii) the Q-independent incoherent background. The elastic neutron diffraction was performed at the lowest and highest temperature at each pressure. Note that the only temperature dependence is the amplitude of the Bragg peak (at 0.21 Å$^{-1}$), which is directly related to the water density. Therefore, we sit at the peak position, measuring the scattering intensity *I(Q=0.21 Å$^{-1}$, T)* as a function of temperature while ramping the temperature from 300 K to 130 K at 0.2 K/min. This ramping rate is slow enough to allow the sample to reach a uniform temperature.

**Data analysis.** In our experiment, we used long wavelength neutrons (λ = 4.7 Å), and focused on the small-angle region (Q from 0.15 Å$^{-1}$ to 0.35 Å$^{-1}$). In such a configuration, neutrons view the water and the silica matrix as continuous media and only the long-range (>18 Å) order is probed. The short-range water-water, silica-silica and water-silica correlation peaks are located at Q values larger than 1.5 Å$^{-1}$, which are beyond the Q range we studied and thus will not concern our measurements. In a small angle diffraction experiment, the neutron scattering intensity distribution *I(Q)* is given by $I(Q) = nV_p^2(\Delta\rho_{sld})^2 \overline{P}(Q)S(Q)$, where n is the number of scattering units (water cylinders) per unit volume, $V_p$ the volume of the scattering unit, $\Delta\rho_{sld} = \rho_{sld}^{D2O} - \rho_{sld}^{MCM}$ the difference of scattering length density (SLD) between the scattering unit $\rho_{sld}^{D2O}$ and the environment $\rho_{sld}^{MCM}$, $\overline{P}(Q)$ the normalized particle structure factor (or form factor) of the scattering unit, and *S(Q)* the inter-cylinder structure factor of a 2-D hexagonal lattice [69]. The SLD of the scattering unit $\rho_{sld}^{D2O}$ is proportional to its mass density $\rho_m^{D2O}$ as $\rho_{sld}^{D2O} = \alpha\rho_m^{D2O}$, where $\alpha = \dfrac{N_A \sum b_i}{M}$, $N_A$ is Avogadro's number, *M* the molecular weight of $D_2O$, $b_i$ the coherent scattering length of the *i*-th atom in the scattering unit. The SLD of the silica material has been determined by a separate contrast matching experiment by hydrating the sample with different ratio of $D_2O$ and $H_2O$. When the molar ratio is [D2O]:[H2O]=0.66:0.34, the Bragg peak is matched out. Compared to water, silica is a rather rigid material. Its thermal expansion coefficient is in the order of 10$^-$

$^6$/K compared to $10^{-3}$/K of water. As shown in Fig.2, the position and the width of the Bragg peaks do not change with temperature, indicating the structure change of the confining matrix is negligible in the measured temperature range. Therefore, based on the above relations, we find that all the variables in the expression for *I(Q)* are independent of temperature except for $\rho_m^{D2O}$. Hence we are able to determine the density of confined $D_2O$ by measuring the temperature-dependent neutron scattering intensity *I(Q)* at the Bragg peak.

The form factor $\overline{P}(Q)$ of a long ($QL>2\pi$) cylinder is given by $\overline{P}(Q) = \pi/QL\left(2J_1(QR)/QR\right)^2$, where *L* and *R* represent the length and the radius of the cylinder respectively, and $J_1(x)$ is the first-order Bessel function of the first kind. The structure factor *S(Q)* can be well approximated by a Lorentzian function. Therefore, the measured neutron scattering intensity is expressed as,

$$I(Q) = nV_p^2\left(\alpha\rho_m^{D2O} - \rho_{sld}^{MCM}\right)^2 \frac{\pi}{QL}\left(\frac{2J_1(QR)}{QR}\right)^2 \left(\frac{\frac{1}{2}\Gamma}{(Q-\frac{2\pi}{d})^2 + (\frac{1}{2}\Gamma)^2}\right) + B\cdot Q^{-\beta} + C, \qquad (1)$$

and at the Bragg peak $Q_0=2\pi/d$,

$$I(Q_0) = A\left(\alpha\rho_m^{D2O} - \rho_{sld}^{MCM}\right)^2 + B\cdot Q_0^{-\beta} + C, \qquad (2)$$

where $\Gamma$ is the FWHM, *C* is the *Q*-independent incoherent background [8]. The approximation of the Bragg peak by a Lorentzian function is purely empirical. The broadening of a diffraction peak comes from many factors, such as the imperfection of the lattice, the instrument resolution, etc. But the choice of the peak formula form will not affect the extraction of the density of water, which merely depends on the peak height.

By fitting the neutron scattering intensity with the above model at the highest and lowest temperature at each pressure, the parameters *B*, *β* and *C* are obtained. We are thus able to subtract the "background" (the second and third term in the above equations) with confidence. We determine the last unknown temperature independent constant *A* by normalizing the density of the highest temperature at each pressure to that of the bulk $D_2O$ taken from NIST Scientific and Technical Database (NIST Chemistry WebBook http://webbook.nist.gov/chemistry/fluid/). When the density of supercooled $D_2O$ is not available (200, 400, 600 bar in Fig.2), the density was normalized to the density of $H_2O$ (at the same temperature and pressure) times 1.109.

Some of the authors have previously used a similar small-angle neutron scattering (SANS) to determine the density of $D_2O$ in MCM-41-S at ambient pressure [8,70]. A similar method has also been used to measure the density of confined toluene [71] and benzene [71]. Recently, the reliability of this method to determine the density of water confined in MCM-41-S has been criticized because of a possible layering effect of water in the pores [72]. However, the scenario hypothesized in [73] assumes the existence of voids in the hydrated pores; this possibility is not consistent with our measurement of a contrast matched sample ($[D_2O]:[H_2O]=0.36:0.34$) in which the diffraction peak is almost completely masked and no evidence of the scattering from the voids can be recognized. The hypothesis of the existence of voids in the hydrated pores originates from a layering density profiles suggested in [73] which implies that water can penetrate into the wall of MCM-41-S. The problem of whether there is void (micropores) on the wall of MCM-41 has been investigated by gas adsorption technique many times since 1993. The great majority concluded that MCM-41 materials do not have any microporosity. Recent experiments suggest that MCM-41 is exclusively mesoporous with no water penetration into the wall [73]. On the basis of these results we believe that the layering density profiles suggested in [74] is unrealistic and inconsistent with the scattering pattern of our measurements.

The effect related to a non-uniform distribution of water in the pore will be contained in the $\overline{P}(Q)$ term. It is generally believed that near the hydrophilic surface such as silica there is a 2-3 Å layer of water with about 10% higher density, while in the center of the pores water distributes uniformly [32,61-67]. When we compare the normalized particle structure factors of this core-shell cylinder and its average, we find that at around the Bragg peak position (Q = 0.2 Å$^{-1}$), the difference of the $\overline{P}(Q)$ is about 5%. When compared to the observed 40% change of the SLD, one realizes that the change of density profile itself would not be enough to contribute to the overall change of the SANS intensities. The small fluctuation at sub-Å scale do not provide further information for density, which is a macroscopic quantity. It should be emphasized that small angle neutron scattering has a "low" spatial resolution and therefore the details of the SLD will have no appreciable effect on the collected data. Therefore, the non-uniform distribution of the water in the pores would account only for a minor correction to the change of the intensity of the Bragg peak, which is mainly influenced by the contrast between the average SLD in the pore and the silica matrix $(\overline{\rho} - \rho_s)^2$. The small angle scattering measurements allow us to determine the average density of fluids in the pores as previously performed by Alba-Simionesco et al [71,72].

**Acknowledgements**

The authors appreciate the efforts of Terrence J. Udovic for his assistance on performing the FANS experiment. The authors are indebted to the in-depth discussions of the content of this paper with H Eugene Stanley, David Chandler, Francesco Sciortino, Giancarlo Franzese and Pablo G Debenedetti. YZ acknowledges the support from the Clifford G. Shull fellowship at ORNL. Research at MIT is supported by Department of Energy Grant DE-FG02-90ER45429, at NTU by Taiwan National Science Council Grant NSC95-2120-M-002-009 and NSC96-2739-M-213-001. This work utilized facilities supported in part by the National Science Foundation under Agreement No. DMR-0454672. We acknowledge the support of the National Institute of Standards and Technology (NIST), U.S. Department of Commerce, in providing the neutron research facilities used in this work. Identification of a commercial product does not imply recommendation or endorsement by the NIST, nor does it imply that the product is necessarily the best for the stated purpose.

**Figures and Figure Captions**

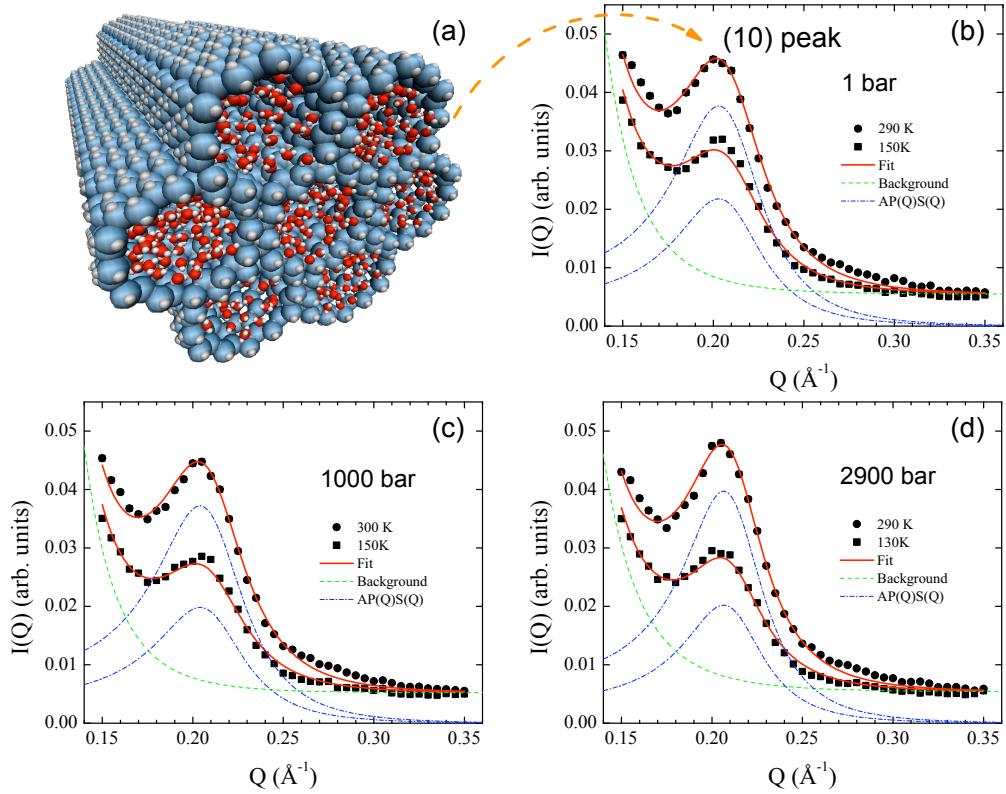

**Figure 1. (a)** Schematic representation of a $D_2O$ hydrated MCM-41-S nanoporous silica crystallite (pore diameter $2R \approx 15$ Å ± 2 Å). **(b-d)** The elastic neutron diffraction intensity $I(Q)$ at three pressures measured by SPINS at NCNR. The structure factor peak at around 0.21 Å$^{-1}$ comes from the (10) plane of the 2-D hexagonal arrangement of the water cylinders in the crystallite. The peak height is proportional to the square of the difference of neutron scattering length density (SLD) between the confined $D_2O$ and the silica matrix, and therefore is a sensitive indicator of the average mass density of $D_2O$ in the pores. By fitting with Eq.(1), the temperature-independent background (green dashed line) and the temperature-dependent elastic diffraction intensities (blue dash-dotted line) can be separated accordingly.

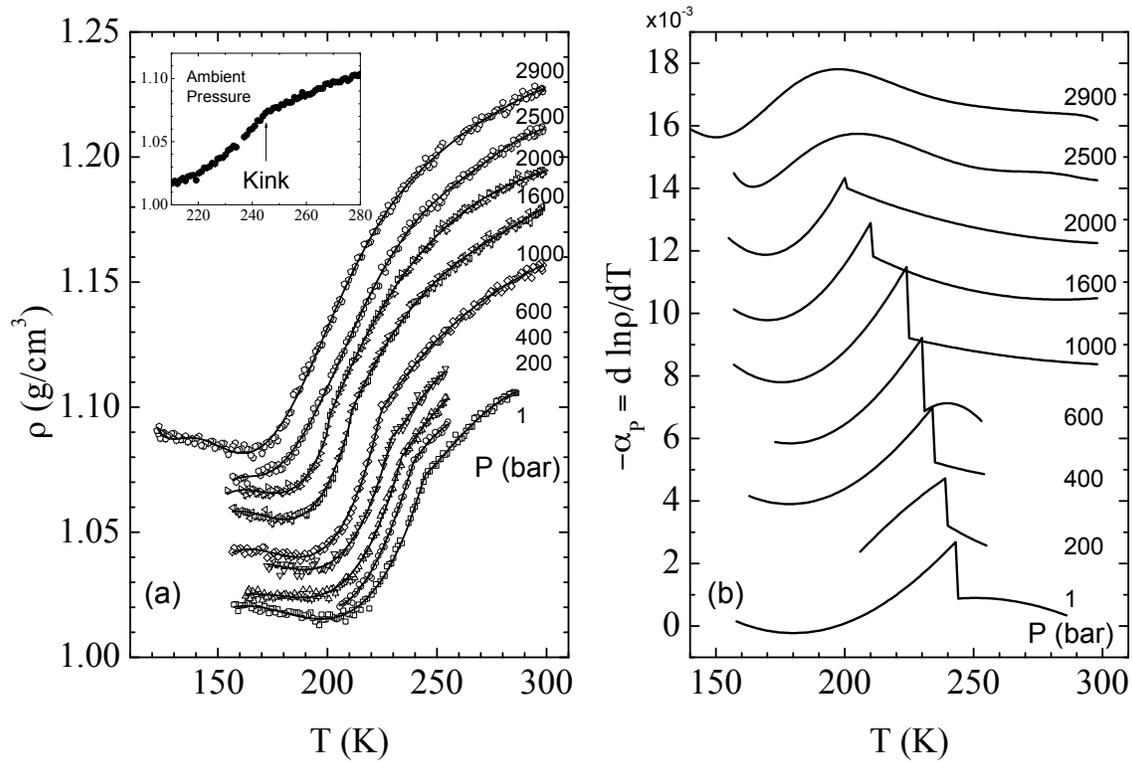

**Figure 2.** (a) Measured isobaric density profiles of confined heavy water by cooling scans. A well-defined "kink" in the density profile can be observed from ambient pressure to ≈1600 bar. The solid lines are fittings of the densities. The inset zooms in the region of the "kink" at ambient pressure. Error bars in the density, due to counting statistics, are smaller than the point size. A larger uncertainty in the density scale is discussed in the text. (b) The isobaric thermal expansion coefficient $\alpha_P$ evaluated from the fitted density curves. The data are shifted by 0.002 between adjacent pressures for clarity. The sudden jump (discontinuity) in $\alpha_P$ below ≈1600 bar directly coming from the "kink" in the density profile is the signature of a 2$^{nd}$-order phase transition by definition. If so, above 2000 bar, the absence of the "kink" in the density profile and the disappearance of the discontinuity in $\alpha_P$ imply the endpoint of the 2$^{nd}$-order transition line.

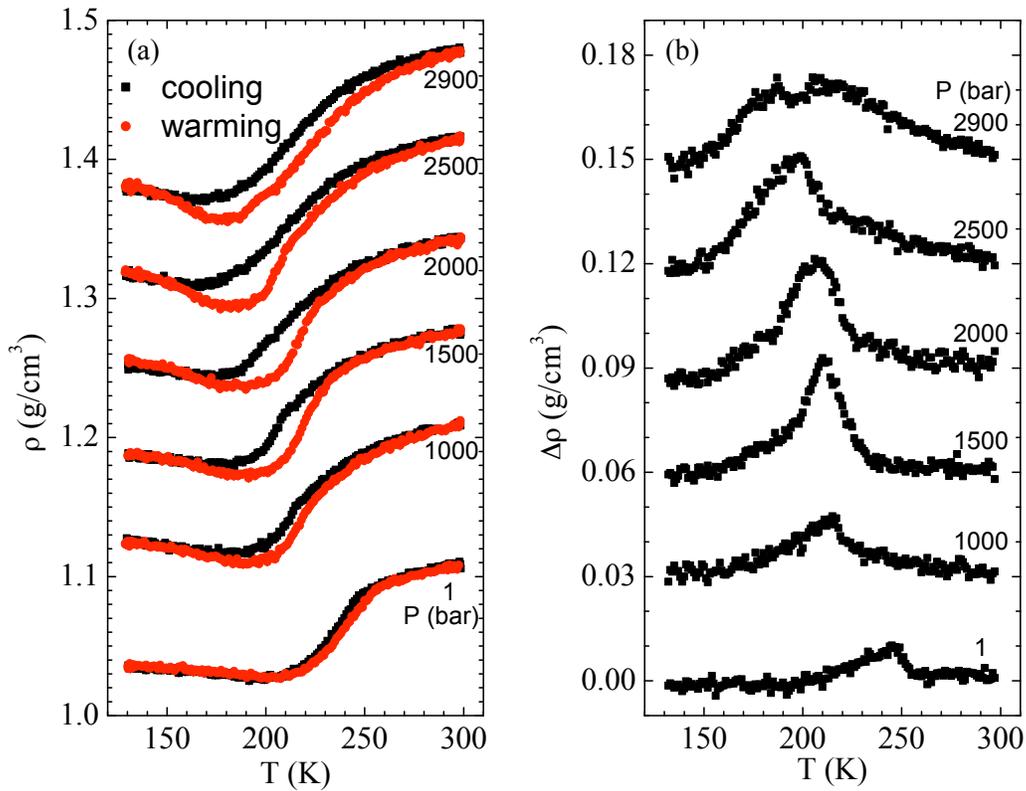

**Figure 3. (a)** Measured isobaric density profiles of confined heavy water by both warming and cooling scans. **(b)** The corresponding density difference between the two scans. The data are shifted by 0.05 and 0.03 g/cm$^3$ between adjacent pressures for clarity in panel **(a)** and **(b)** respectively. A hysteresis phenomenon becomes prominent at ≈1500 bar and above, which is a direct evidence of the phase separation when a 1$^{st}$-order phase transition line is crossed.

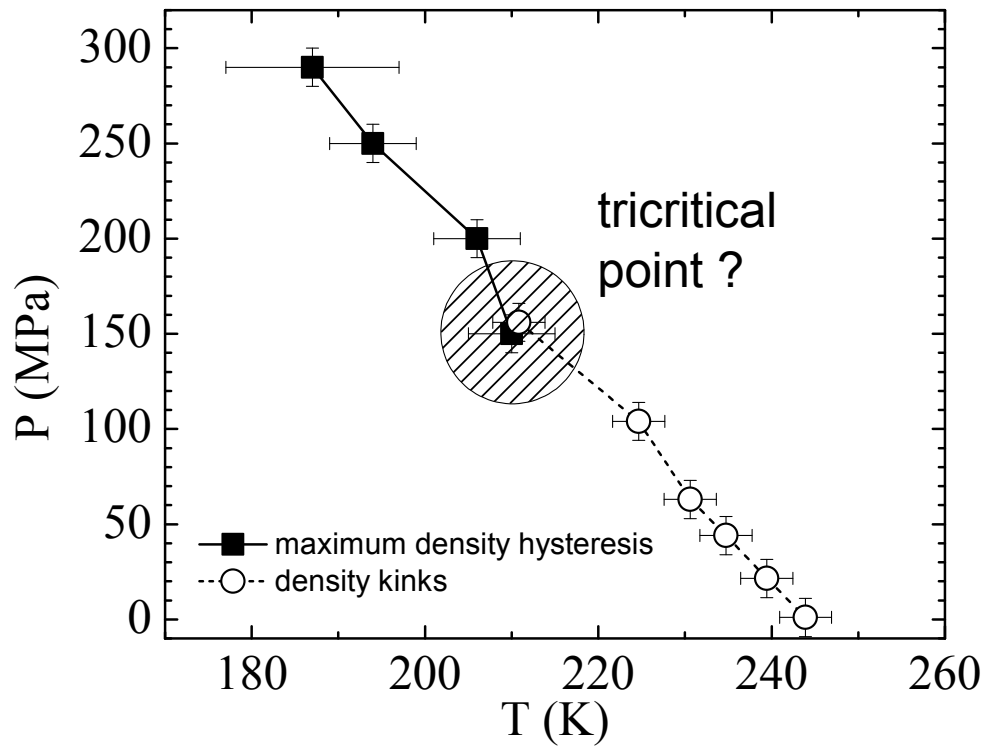

**Figure 4.** P-T phase diagram of the confined heavy water. The empty circles, connected by the dotted line (a possible $2^{nd}$-order transition line), are determined from the position of the "kinks" in the density profiles. The filled squares, connected by the solid line (a possible $1^{st}$-order transition line), are determined from the position of the maximum density hysteresis. Therefore, the shadow area connecting the $2^{nd}$-order and $1^{st}$-order lines is where one may find a "tricritical point" [35].

# References

[1] Ivan Brovchenko and Alla Oleinikova, *Interfacial and Confined Water* (Elsevier, 2008), p. 305.
[2] David Chandler, Nature **437**, 640-7 (2005).
[3] Michael D Fayer and Nancy E Levinger, Annual Review Of Analytical Chemistry **3**, 89-107 (2010).
[4] Pablo G. Debenedetti and H. Eugene Stanley, Physics Today **56**, 40-46 (2003).
[5] O Mishima and H. Eugene Stanley, Nature **396**, 329-335 (1998).
[6] Pablo G. Debenedetti, Journal Of Physics: Condensed Matter **15**, R1669-R1726 (2003).
[7] C. A. Angell, Annual Review Of Physical Chemistry **34**, 593-630 (1983).
[8] D. Liu, Y. Zhang, C.-C. Chen, C.-Y. Mou, P. H. Poole, and S.-H. Chen, Proceedings Of the National Academy Of Sciences Of the United States Of America **104**, 9570-9574 (2007).
[9] Francesco Mallamace, Caterina Branca, Matteo Broccio, Carmelo Corsaro, Chung-Yuan Mou, and Sow-Hsin Chen, Proceedings Of the National Academy Of Sciences Of the United States Of America **104**, 18387-91 (2007).
[10] Yang Zhang, Kao-Hsiang Liu, Marco Lagi, Dazhi Liu, Kenneth C Littrell, Chung-Yuan Mou, and Sow-Hsin Chen, The Journal Of Physical Chemistry B **113**, 5007-10 (2009).
[11] C. A. Angell, MRS Bulletin **33**, 544-555 (2008).
[12] S. Sastry, Pablo G. Debenedetti, F. Sciortino, and H. Eugene Stanley, Physical Review E **53**, 6144-6154 (1996).
[13] H. Eugene Stanley and J. Teixeira, The Journal Of Chemical Physics **73**, 3404-3422 (1980).
[14] P. H. Poole, F. Sciortino, U. Essmann, and H. Eugene Stanley, Nature **360**, 324-328 (1992).
[15] Dietmar Paschek, Physical Review Letters **94**, 217802 (2005).
[16] C Austen Angell, Science **319**, 582-7 (2008).
[17] O Mishima, L. D. Calvert, and E. Whalley, Nature **314**, 76-78 (1985).
[18] O Mishima, Physical Review Letters **85**, 334-336 (2000).
[19] Osamu Mishima and Yoshiharu Suzuki, Nature **419**, 599-603 (2002).
[20] S. Klotz, Th. Strässle, R. Nelmes, J. Loveday, G. Hamel, G. Rousse, B. Canny, J. Chervin, and A. Saitta, Physical Review Letters **94**, 025506 (2005).
[21] O Mishima and H. Eugene Stanley, Nature **392**, 164-168 (1998).
[22] A. K. Soper and M. A. Ricci, Physical Review Letters **84**, 2881-2884 (2000).
[23] T. Tokushima, Y. Harada, O. Takahashi, Y. Senba, H. Ohashi, L.G.M. Pettersson, A. Nilsson, and S. Shin, Chemical Physics Letters **460**, 387-400 (2008).
[24] C Huang, K T Wikfeldt, T Tokushima, D Nordlund, Y Harada, U Bergmann, M Niebuhr, T M Weiss, Y Horikawa, M Leetmaa, M P Ljungberg, O Takahashi, A Lenz, L Ojamäe, A P Lyubartsev, S Shin, L G M Pettersson, and A Nilsson, Proceedings Of the National Academy Of Sciences Of the United States Of America **106**, 15214-8 (2009).
[25] Francesco Mallamace, Matteo Broccio, Carmelo Corsaro, Antonio Faraone, Domenico Majolino, Valentina Venuti, Li Liu, Chung-Yuan Mou, and Sow-Hsin Chen, Proceedings Of the National Academy Of Sciences Of the United States Of America **104**, 424-8 (2007).
[26] Chae Un Kim, Buz Barstow, Mark W Tate, and Sol M Gruner, Proceedings Of the National Academy Of Sciences Of the United States Of America **106**, 4596-600 (2009).
[27] D Banerjee, S N Bhat, S V Bhat, and D Leporini, Proceedings Of the National Academy Of Sciences Of the United States Of America **106**, 11448-53 (2009).
[28] C. A. Tulk, C. J. Benmore, J. Urquidi, D. D. Klug, J. Neuefeind, B. Tomberli, and P. A. Egelstaff, Science **297**, 1320-1323 (2002).
[29] Gary N I Clark, Greg L Hura, Jose Teixeira, Alan K Soper, and Teresa Head-Gordon, Proceedings Of the National Academy Of Sciences Of the United States Of America **107**, 14003-14007 (2010).
[30] A. K. Soper, Molecular Physics **106**, 2053-2076 (2008).

[31] Ateeque Malani, K G Ayappa, and Sohail Murad, The Journal Of Physical Chemistry B **113**, 13825-39 (2009).
[32] P. Gallo, M. Rovere, and S.-H. Chen, The Journal Of Physical Chemistry Letters **1**, 729-733 (2010).
[33] Peter H Poole, Ivan Saika-Voivod, and Francesco Sciortino, Journal Of Physics: Condensed Matter **17**, L431-L437 (2005).
[34] L D Landau and E.M. Lifshitz, *Statistical Physics, Part 1* (Butterworth-Heinemann, 1980), p. 544.
[35] J Niemela and R J Donnelly, Journal Of Low Temperature Physics **98**, 1-16 (1995).
[36] J A Lipa, D R Swanson, J A Nissen, and T C P Chui, Physica B **197**, 239-248 (1994).
[37] Masaharu Oguni, Satoshi Maruyama, Kenji Wakabayashi, and Atsushi Nagoe, Chemistry, an Asian Journal **2**, 514-20 (2007).
[38] Giancarlo Franzese and H Eugene Stanley, Journal Of Physics: Condensed Matter **19**, 205126 (2007).
[39] Marco G. Mazza, Kevin Stokely, H. Eugene Stanley, and Giancarlo Franzese, arXiv:0807.4267 (2008).
[40] Limei Xu, Pradeep Kumar, S V Buldyrev, S-H Chen, P H Poole, F Sciortino, and H E Stanley, Proceedings Of the National Academy Of Sciences Of the United States Of America **102**, 16558-62 (2005).
[41] P. H. Poole, F. Sciortino, U. Essmann, and H. Eugene Stanley, Physical Review E **48**, 3799-3817 (1993).
[42] L. Gelb, K. Gubbins, R. Radhakrishnan, and M. Sliwinska-Bartkowiak, Reports On Progress In Physics **62**, 1573-1659 (1999).
[43] R. Evans, Journal Of Physics: Condensed Matter **2**, 8989-9007 (1990).
[44] Gerhard H Findenegg, Susanne Jähnert, Dilek Akcakayiran, and Andreas Schreiber, Chemphyschem : A European Journal Of Chemical Physics and Physical Chemistry **9**, 2651-9 (2008).
[45] K. Morishige and K. Kawano, The Journal Of Chemical Physics **110**, 4867-4872 (1999).
[46] Daisuke Takaiwa, Itaru Hatano, Kenichiro Koga, and Hideki Tanaka, Proceedings Of the National Academy Of Sciences Of the United States Of America **105**, 39-43 (2008).
[47] C. A. Angell, Chemical Reviews **102**, 2627-2650 (2002).
[48] G P Johari, Andreas Hallbrucker, and Erwin Mayer, Nature **330**, 552-553 (1987).
[49] V Velikov, S Borick, and C A Angell, Science **294**, 2335-2338 (2001).
[50] C Austen Angell, Annual Review Of Physical Chemistry **55**, 559-83 (2004).
[51] J Swenson, Journal Of Physics: Condensed Matter **16**, S5317-S5327 (2004).
[52] A Faraone, L Liu, C-Y Mou, C-W Yen, and S-H Chen, The Journal Of Chemical Physics **121**, 10843-6 (2004).
[53] Li Liu, Sow-Hsin Chen, Antonio Faraone, Chun-Wan Yen, and Chung-Yuan Mou, Physical Review Letters **95**, 117802 (2005).
[54] C Domb and J L Lebowitz, *Phase Transitions and Critical Phenomena, Vol. 9* (Academic Press, 1984).
[55] P. M. Chaikin and T. C. Lubensky, *Principles Of Condensed Matter Physics* (Cambridge University Press, 2000), p. 720.
[56] B Widom, Journal Of Physical Chemistry **100**, 13190-13199 (1996).
[57] E H Graf, D M Lee, and J D Reppy, Physical Review Letters **19**, 417-& (1967).
[58] E Cohen, Science **197**, 11-16 (1977).
[59] R J Birgeneau, G Shirane, M Blume, and W C Koehler, Physical Review Letters **33**, 1098-1101 (1974).
[60] Yu Liu, Wenzhong Zhang, and Thomas J. Pinnavaia, Journal Of the American Chemical Society **122**, 8791-8792 (2000).
[61] Shuangyan Xu, George W Scherer, T S Mahadevan, and Stephen H Garofalini, Langmuir : The ACS Journal Of Surfaces and Colloids **25**, 5076-83 (2009).

[62] P. Gallo, M. A. Ricci, and M. Rovere, The Journal Of Chemical Physics **116**, 342-346 (2002).
[63] C. Hartnig, W. Witschel, E. Spohr, P. Gallo, M. A. Ricci, and M. Rovere, Jouranl Of Molecular Liquids **85**, 127-137 (2000).
[64] Nicolas Giovambattista, Peter J Rossky, and Pablo G Debenedetti, The Journal Of Physical Chemistry B **113**, 13723-34 (2009).
[65] S. H. Lee and P. J. Rossky, The Journal Of Chemical Physics **100**, 3334-3345 (1994).
[66] Amish J Patel, Patrick Varilly, and David Chandler, The Journal Of Physical Chemistry B **114**, 1632-7 (2010).
[67] Ezequiel de la Llave, Valeria Molinero, and Damián A Scherlis, The Journal Of Chemical Physics **133**, 034513 (2010).
[68] Antonio Faraone, Kao-Hsiang Liu, Chung-Yuan Mou, Yang Zhang, and Sow-Hsin Chen, The Journal Of Chemical Physics **130**, 134512 (2009).
[69] Sow-Hsin Chen, Annual Review Of Physical Chemistry **37**, 351-399 (1986).
[70] Dazhi Liu, Yang Zhang, Yun Liu, Jianlan Wu, Chia-Cheng Chen, Chung-Yuan Mou, and Sow-Hsin Chen, The Journal Of Physical Chemistry B **112**, 4309-12 (2008).
[71] Denis Morineau, Yongde Xia, and Christiane Alba-Simionesco, The Journal Of Chemical Physics **117**, 8966 (2002).
[72] Yongde Xia, Gilberte Dosseh, Denis Morineau, and Christiane Alba-Simionesco, The Journal Of Physical Chemistry B **110**, 19735-44 (2006).
[73] R. Mancinelli, F. Bruni, and M. A. Ricci, The Journal Of Physical Chemistry Letters **1**, 1277-1282 (2010).
[74] A Silvestre-Albero, E O Jardim, E Bruijn, V Meynen, P Cool, A Sepúlveda-Escribano, J Silvestre-Albero, and F Rodríguez-Reinoso, Langmuir : The ACS Journal Of Surfaces and Colloids **25**, 939-943 (2008).